# High-pressure x-ray diffraction study on the structure and phase transitions of the defect-stannite $ZnGa_2Se_4$ and defect-chalcopyrite $CdGa_2S_4$

D. Errandonea[1,*], Ravhi S. Kumar[2], F. J. Manjón[3], V. V. Ursaki[4], and I. M. Tiginyanu[4]

[1]MALTA Consolider Team, Departamento de Física Aplicada-ICMUV, Fundación General de la Universitad de Valencia, Edificio de Investigación, c/Dr. Moliner 50, 46100 Burjassot (Valencia), Spain

[2]High Pressure Science and Engineering Center, Department of Physics and Astronomy, University of Nevada Las Vegas, 4505 Maryland Parkway, Las Vegas, Nevada 89154-4002, USA

[3]MALTA Consolider Team, Departamento de Física Aplicada-IDF, Universitat Politècnica de València, Cno. de Vera s/n, 46022 València, Spain

[4]Institute of Applied Physics, Academy of Sciences of Moldova, 2028 Chisinau, Moldova

**Abstract:** X-ray diffraction measurements on the sphalerite-derivatives $ZnGa_2Se_4$ and $CdGa_2S_4$ have been performed upon compression up to 23 GPa in a diamond-anvil cell. $ZnGa_2Se_4$ exhibits a defect tetragonal stannite-type structure ($I\bar{4}2m$) up to 15.5 GPa and in the range from 15.5 GPa to 18.5 GPa the low-pressure phase coexists with a high-pressure phase, which remains stable up to 23 GPa. In $CdGa_2S_4$, we find the defect

* Corresponding author, Email: daniel.errandonea@uv.es, Fax: (34) 96 3543146, Tel.: (34) 96 354 4475

tetragonal chalcopyrite-type structure ( $I\bar{4}$ ) is stable up to 17 GPa. Beyond this pressure a pressure-induced phase transition takes place. In both materials, the high-pressure phase has been characterized as a defect-cubic NaCl-type structure ( $Fm\bar{3}m$ ). The occurrence of the pressure induced phase transitions is apparently related with an increase of the cation disorder on the semiconductors investigated. In addition, the results allow the evaluation of the axial compressibility and the determination of the equation of state for each compound. The obtained results are compared with those previously reported for isomorphic digallium sellenides. Finally, a systematic study of the pressure-induced phase transition in twenty-three different sphalerite-related $ABX_2$ and $AB_2X_4$ compounds indicates that the transition pressure increases as the ratio of the cationic radii and anionic radii of the compounds increases.

## I. Introduction

Zinc digallium selenide ($ZnGa_2Se_4$) and cadmium digallium sulphide ($CdGa_2S_4$) are tetrahedrally coordinated $A^{II}B_2^{III}X_4^{VI}$ defective compounds the structure of which is still contradictory discussed in the literature. While some studies suggest a defect-chalcopyrite structure ($I\bar{4}$) others report a defect-stannite structure ($I\bar{4}2m$) for these compounds. Both structures are tetragonal and structurally related to the cubic sphalerite structure ($F\bar{4}3m$), commonly known as zinc-blende, with only differences arising due to slightly deviations on the atomic positions of the anions. This family of semiconductors are of interest as possible infrared-transmitting windows materials. They are also applied in various nonlinear optical devices and as gyrotropic media in narrow-band optical filters. In addition, these compounds are promising optoelectronic materials due to their high values of nonlinear susceptibility, optical activity, intense luminescence, and high photosensitivity. Some compounds like $CdGa_2Se_4$ and $CdAl_2S_4$ have already found practical applications as tunable filters and ultraviolet photodetectors [1, 2]. High-pressure studies on $A^{II}B_2^{III}X_4^{VI}$ compounds are receiving increasing interest in the last years. In particular, these materials have been extensively studied by Raman spectroscopy [3 – 6] because of pressure-induced phase transitions. However, the structure of the high-pressure phases has been so far characterized only for two compounds, $CdGa_2Se_4$ [7] and $MnGa_2Se_4$ [8]. In both materials the high-pressure phase has been determined as a cubic NaCl-type structure.

In the present work, we report angle dispersive x-ray diffraction (ADXRD) measurements performed on $ZnGa_2Se_4$ and $CdGa_2S_4$ as a function of pressure in a diamond-anvil cell (DAC) at room temperature up to 23 GPa. From these experiments, we have determined the effect of pressure on the lattice parameters and atomic positions, as well as observed pressure-driven structural phase transitions. In both compounds, the high-pressure phase has been characterized and assigned to a defect NaCl-type structure. We also determined the equation of state (EOS) of $CdGa_2S_4$ and $ZnGa_2Se_4$ and further discuss the systematics of pressure-induced phase transitions on sphalerite derivatives.

## II. Experiment

Single crystals of $CdGa_2S_4$ and $ZnGa_2Se_4$ were grown by chemical vapor method using iodine as a transport agent. The as grown crystals represent triangular prisms with mirror surfaces. Chemical and structural analyses have shown the stoichiometric composition of the crystals and no spurious phases were observed. ADXRD experiments were carried out at room temperature under compression up to 23 GPa using a DAC at Sector 16-IDB of the HPCAT, at the Advanced Photon Source (APS). $CdGa_2S_4$ was studied with an incident monochromatic wavelength of 0.36806 Å and $ZnGa_2Se_4$ with a wavelength of 0.41521 Å. The samples used in the experiments were pre-pressed pellets prepared using a finely ground powder obtained from the as grown single crystals. These pellets were loaded in a 130 μm hole of a rhenium gasket in a Mao-Bell-type DAC with diamond-culet sizes of 350 μm. A few ruby grains were also loaded with the sample for pressure determination [9] and silicone oil was used as pressure-transmitting medium [10, 11]. The monochromatic x-ray beam was focused down to $10 \times 10$ $\mu m^2$ using Kickpatrick-Baez mirrors. The images were collected using a MAR345 image plate located 380 mm away from the sample in the $CdGa_2S_4$ experiment and 300 mm away from the sample in the $ZnGa_2Se_4$ experiment. The collected images were integrated and corrected for distortions using FIT2D [12]. The structure refinements were performed using the POWDERCELL [13] program package.

## III. Results and discussion

### III.A. The structure of $CdGa_2S_4$ and $ZnGa_2Se_4$

$ZnGa_2Se_4$ has been reported to have either a tetragonal defect-chalcopyrite ($I\bar{4}$, Z = 2) or a tetragonal defect-stannite ($I\bar{4}2m$, Z = 2) structure [14]. Both structures are shown in Fig. 1. Recent Raman measurements supported the space group $I\bar{4}$ for $ZnGa_2Se_4$ [15], but accurate neutron diffraction studies unambiguously established that it belongs to the space group $I\bar{4}2m$ [16]. Our experiments agree with the conclusions drawn from the neutron experiments. After a Rietveld refinement of an x-ray diffraction pattern collected at ambient pressure (0.1 MPa) outside the DAC the following structural parameters for defect-stannite $ZnGa_2Se_4$ were obtained: $a$ = 5.512(3) Å and $c$ = 10.963(6) Å. The residuals of the refinement are $R_F^2$ = 2.52 %, $R_{WP}$= 3.97 %, $R_P$ = 2.07 %. The atomic positions obtained for the defect-stannite structure are summarized in Table I. According to the site occupation fraction obtained in the structural refinement, a partial cation order is present in $ZnGa_2Se_4$ [16]. However, since Ga and Zn have nearly equal x-ray scattering factors, it is hard to distinguish between this model with others considering a complete cation order (Zn at 2a and Ga at 4d) or a complete cation disorder (Zn:Ga =1/3:2/3 at 2a and 4d). Indeed these models resulted in slightly larger R-factors if considered for the structural refinement than the partial ordering model (see atomic positions in Table I). Nevertheless, based upon this fact and the conclusions drawn by Hanada [16], we think it can be concluded that the structure of $ZnGa_2Se_4$ is defect-stannite as summarized in Table I. This type of structure is observed in the minerals famatinite ($Cu_3SbS_4$) and stannite ($Cu_2FeSnS_4$) [17]. According to the partial cation ordering model, in $ZnGa_2Se_4$, half of the Ga atoms occupy the position of Fe in stannite, the other half and the Zn atoms occupy the position of Cu and the Se atoms are located at the position of S. The structure of $ZnGa_2Se_4$ is called defect-stannite (some authors call it defect-famatinite) because the Sn atoms (located at 2b in stannite) are replaced by vacancies in this compound. Further, in $ZnGa_2Se_4$ the Zn and Ga cations and the vacancies are in tetrahedral coordination, but these tetrahedra differ in dimension and bond angles. For the Ga atoms located at 2a, the Ga-Se bond distance is 2.42 Å and for the Ga and Zn atoms located at 4d, the Ga(Zn)-Se bond distance is 2.44 Å. The vacancy-Se distance is 2.24 Å. It is important to note that the cation-anion bonds are close to those found in ZnSe (2.45 Å) [18] and in the high-temperature phase of $Ga_2Se_3$ (2.42 Å) [19]. However, the vacancy-anion distance is much shorter, due to the fact that the Se atoms are displaced from the ideal (1/4, 1/4, 1/8) towards the vacancy.

Regarding $CdGa_2S_4$, it is accepted that it has a defect-chalcopyrite structure ($I\bar{4}$) in which one cation site is also vacant (at 2d) [20]. In a similar way to $ZnGa_2Se_4$, we also analyzed $CdGa_2S_4$ at ambient conditions outside the DAC. Rietveld refinement of the x-ray diffraction pattern collected at ambient conditions provided the following structural parameters for the defect-chalcopyrite structure of $CdGa_2S_4$: $a$ = 5.536(3) Å and $c$ = 10.160(6) Å with refinement residuals $R_F^2$ = 2.02 %, $R_{WP}$= 2.87 %, $R_P$ = 1.67 %. The lattice parameters obtained agree with the values available in the literature. The atomic positions determined in the structural refinement are summarized in Table II. The structure observed for $CdGa_2S_4$ is typical to the mineral kesterite ($Cu_2ZnSnS_4$) [17]. In the case of $CdGa_2S_4$, half of the Ga atoms occupy the position of Sn as in kesterite, the other half and the Cd atoms occupy the position of Cu, while the S atoms stay at the same position. In this case, the vacancies occupy the position of Zn. The structure of $CdGa_2S_4$ is called defect-chalcopyrite because it can be constructed from the chalcopyrite structure of $CuGaS_2$ ($I\bar{4}2d$) [21] by doubling the formula unit and replacing Cu by Cd and a vacancy to maintain the valence. In $CdGa_2S_4$ the Cd and Ga cations and the vacancies are in tetrahedral coordination, but these tetrahedra differ in dimension and bond angles. We found the Cd-S bond distance to be 2.52 Å, the Ga-S distance to be 2.33 Å (for the Ga atoms at 2b) and 2.29 Å (for Ga atoms at 2c). The vacancy-S distance was found to be 2.22 Å. The Cd-S distance is exactly the same as in CdS [22] and the Ga-S distances are close to the ones reported for $Ga_2S_3$ (2.22 Å) [23].

**III.B. Pressure-induced phase transitions**

A summary of the results obtained in the high-pressure x-ray diffraction experiments for $ZnGa_2Se_4$ and $CdGa_2S_4$ is shown in Figs. 2 and 3, respectively. In $ZnGa_2Se_4$, we did not observe any substantial change in the x-ray diffraction patterns up to 10.37 GPa. At this pressure we found that most of the diffraction peaks become broader. The peak broadening increases with pressure beyond 10.37 GPa. This broadening can be clearly seen in Fig. 2 looking at the peak located near $2\theta = 7.6^o$ - the (112) reflection – whose full width at half maximum is smaller than 0.2º below 10.37 GPa, but it becomes 0.3º at 10.37 GPa and 0.4º at 17.5 GPa. A similar phenomenon has been observed around 10 GPa in Raman measurements [3, 5]. Peak broadening in both cases could be related with an increase of the crystalline disorder induced by pressure. It could be also related to non-hydrostatic effects due to the use of silicone oil as pressure-transmitting medium [11]. These effects cannot be neglected beyond 10 GPa. However, as we will show latter, the observed reduction of the anion distortion parameter $\sigma_0$ under compression supports the first hypothesis. At 15.5 GPa we observe the appearance of new diffraction peaks. The intensity of these peaks is found to increase with pressure. They are indicated by asterisks in the pattern collected at 17.5 GPa. At 18.5, the intensity of the peaks assigned to the low-pressure phase decreased considerably and the new peaks became dominant in the diffraction pattern. The peaks corresponding to the low-pressure phase disappeared at 19.2 GPa. All these changes indicated the occurrence of a pressure-driven phase transition. We have located the onset of the transition at 15.5 GPa and a coexistence of the low- and high-pressure phases was noticed up to 18.5 GPa. The transition is complete at 19.2 GPa. Upon further compression the high-pressure phase appears to remain stable up to 23 GPa, the highest pressure reached in our experiments. On pressure release, we reduced the pressure in three steps; from 23 GPa to 14.7 GPa, 2 GPa, and ambient pressure, respectively. Apparently the phase transition is irreversible since the diffraction pattern, collected on the sample recovered at ambient pressure, resembled very much the diffraction patterns of the high-pressure phase. In Raman experiments, decrease in the intensity of Raman signal has been detected beyond 14.7 GPa and a total disappearance at 18.9 GPa [3, 5]. On decompression the Raman signal was not completely recovered, as this fact is attributed to a partial amorphization of $ZnGa_2Se_4$. The pressure-induced changes in the Raman spectra are consistent with the changes that we observed in the x-ray diffraction pattern. In both cases the onset of a phase transition is detected around 15 GPa with a coexistence of the low- and high-pressure phases up to around 19 GPa. In $CdGa_2S_4$, we did not observe any substantial change in the diffraction patterns up to 17 GPa. Beyond this pressure, changes in the patterns occur suggesting also the appearance of a high-pressure phase. At 21 GPa, only the high-pressure phase is present suggesting that the phase transition has been completed. We have identified the phase transition pressure as 19(2) GPa. Raman measurements reported for $CdGa_2S_4$ show that the Raman signal irreversibly disappears at 15 GPa [3]. This fact was attributed to a phase transition to a NaCl-type structure and we believe that the transition we observed at slightly higher pressures is also to a cubic phase. In contrast with $ZnGa_2Se_4$, we did not collect diffraction patterns upon pressure release for $CdGa_2S_4$, so we cannot give any information on the reversibility of the reported phase transition.

We performed full profile Rietveld refinement of the diffraction patterns collected for the high-pressure phase. Similar to the defect-chalcopyrite selenides $MnGa_2Se_4$ and $CdGa_2Se_4$ [7, 8], the high-pressure phases of $ZnGa_2Se_4$ and $CdGa_2S_4$ have a higher-symmetry cubic NaCl-type structure ($Fm\bar{3}m$, Z = 1), where the Se atoms are located at the 4b site, while the Zn, Ga, and the vacancies are located at the 4a sites. The phase transition implies an increase in the symmetry of the crystals and is accompanied by a change in the coordination of the cations from tetrahedral to octahedral. The similitude between the high-pressure behavior of defect-chalcopyrite and defect-stannite digallium tetraselenides and tetrasulphides is not surprising since both kinds of structures are closely related. As a matter of fact, the *ideal* chalcopyrite and stannite structures are themselves ordered (tetragonal) versions of sphalerite (zinc-blende ZnSe) with the unit cell nearly doubled along [001] [24]. Both structures can be thought as superstructures of ZnSe (or ZnS) with the Zn atoms being replaced alternatively by Ga and Cd (Mn or Zn) atoms and vacancies. The only difference between them is the movement of the Se (S) atom away from (x,x,z) in $I\bar{4}2m$ to (x,y,z) in $I\bar{4}$, reducing the site symmetry around the vacancy and one cation site from $\bar{4}2m$ to $\bar{4}$. ZnSe, ZnS, and isostructural tetrahedrally-coordinated semiconductors undergo zinc-blende to rock-salt pressure-induced phase transitions [25]. Therefore, based upon crystallochemical arguments a transition to a NaCl-type structure is also expected in both tetragonal tetraselenides and tetrasulphides [26]. It is worth mentioning that, x-ray diffraction experiments performed on $MnGa_2Se_4$ also reported an irreversible NaCl-type transition very similar to our observation on $ZnGa_2Se_4$ [8]. On the other hand, x-ray diffraction experiments for $CdGa_2Se_4$ show that the NaCl-type phase transforms into a zinc-blende-type structure below 5 GPa upon decompression and this structure is different to the defect-chalcopyrite structure [7]. Our results are in good agreement with the studies of Marquina *et al.* [8], and also they agree well with the conclusions drawn from Raman measurements on $CdGa_2Se_4$, $CdGa_2S_4$, $ZnGa_2Se_4$, and $ZnGa_2S_4$ [3, 6]. In these studies the samples recovered at ambient pressure were found to be Raman inactive, while the zinc-blende structure is expected to have two Raman active phonons. In contrast with these results, Raman studies performed by Mitani *et al.* [6] showed that Raman bands were recovered for $CdGa_2Se_4$ upon decompression, but they do not correspond to the defect-chalcopyrite structure. On top of that, unpublished optical-absorption measurements on single-crystalline $CdGa_2Se_4$ and $ZnGa_2Se_4$ show a new phase on pressure release from the high-pressure phase [27]. In order to solve this puzzle, clearly more research is needed to clarify whether a metastable phase is obtained upon pressure release in digallium tetraselenides and tetrasulphides or not.

### III.C. Pressure dependence of the lattice parameters and equations of state

From the refinement of x-ray diffraction patterns we have obtained the pressure dependence of the lattice parameters for the low- and high-pressure phases. The pressure evolution of the structural parameters and the atomic volume ($V$) of $ZnGa_2Se_4$ and $CdGa_2S_4$ are shown in Figs. 4 and 5, respectively. To make it easier for comparison between the low- and high-pressure phases, we plotted $2V$ instead of $V$ for the NaCl-type phase. In $ZnGa_2Se_4$, the compression of the low-pressure phase is slightly anisotropic up to 10.4 GPa. The axial ratio $c/a$ increases from 1.988 at ambient pressure to 2 at 10.4 GPa and beyond this pressure the $c/a$ remains constant within the uncertainty of the experiments. In $CdGa_2S_4$, the compression of the low-pressure phase is highly anisotropic. In particular, $c/a$ increases from 1.835 to 1.913 from ambient pressure to 17 GPa following a nearly linear pressure dependence. The larger increase of the axial ratio in $CdGa_2S_4$ is related to the smaller compressibility of the $c$-axis in this compound. A similar behavior has been observed in $MnGa_2Se_4$ and $CdGa_2Se_4$; i.e.

apparently the four sphalerite-derivative compounds become more symmetric prior to the occurrence of the phase transition. In $ZnGa_2Se_4$, the lattice parameters at 18.5 GPa are $a$ = 5.064 Å and $c$ = 10.156 Å for the low-pressure phase and $a$ = 5.064 Å for the high-pressure phase. Therefore, a volume collapse of about 4.6 % is observed at the phase transition. In $CdGa_2S_4$, the lattice parameters at 17 GPa are $a$ = 5.064 Å and $c$ = 10.156 Å for the low-pressure phase, while for the high-pressure phase we obtained $a$ = 4.911 Å at 22 GPa, which implies a volume collapse of about 5 %. The presence of the volume collapses is in agreement with the results reported for $MnGa_2Se_4$ and $CdGa_2Se_4$ [7, 8]. It also indicates that the reported transition is a first-order transition.

The pressure-volume curves shown in Figs. 4 and 5 were analyzed using a Birch-Murnaghan EOS: $P = \frac{3}{2} B_0 \left(x^{7/3} - x^{5/3}\right)\left[1 + \frac{3}{4}\left(B_0' - 4\right)\left(x^{2/3} - 1\right)\right]$, with $x = V_0/V$, where the parameters $V_0$ , $B_0$, and $B_0'$ are the zero-pressure volume, bulk modulus, and pressure derivative of the bulk modulus, respectively. For the defect-stannite phase of $ZnGa_2Se_4$, by fixing $V_0$ = 333.08 Å$^3$ (the measured value at ambient pressure outside the DAC), we obtained $B_0$ = 47(2) GPa and $B_0'$ = 3.9(3). For the NaCl-type structure of $ZnGa_2Se_4$, by fixing $V_0$ = 156.5 Å$^3$ (the measured value at ambient pressure in the recovered sample) and $B_0'$ = 4, we obtained $B_0$ = 50(2) GPa. For the defect-chalcopyrite phase of $CdGa_2S_4$, by fixing $V_0$ = 311.38 Å$^3$ (the measured value at ambient pressure outside the DAC), we obtained $B_0$ = 64(2) GPa and $B_0'$ = 4.1(3). From these results we conclude that the low-pressure phase of $ZnGa_2Se_4$ has a similar compressibility to the low-pressure phases of $MnGa_2Se_4$, $B_0$ = 44(2) GPa [8] and $CdGa_2Se_4$, $B_0$ = 42(2) GPa [7] for the low-pressure phase. For the high-pressure phase of $CdGa_2S_4$ we do not have enough data points to determine its EOS parameters. However, apparently the cubic high-pressure phase is less compressible than the tetragonal phases, as observed by us in $ZnGa_2Se_4$ and by Marquina *et al.* in $MnGa_2Se_4$ [8]. We also conclude that the tetraselenides are more compressible than the tetrasulphides, which is consistent with the fact that for binary compounds like ZnSe, ZnS, CdSe, and CdS, the sulphides are harder than the selenides [25]. Previously a bulk modulus of 66 GPa (88 GPa) was calculated for $ZnGa_2Se_4$ ($CdGa_2S_4$) using the equation deduced by Baranovsky within a tight-binding approach [28]. The same approach predicts a bulk modulus of 64 GPa for $CdGa_2Se_4$. Our and previous measurements show that the tight-binding model overestimates $B_0$ by more than 40%.

We also analyzed the pressure evolution of bond distances. From our Rietveld refinements we found that not only the structure of $ZnGa_2Se_4$ becomes more symmetric under compression, but also the position of the Se atoms gradually approach the ideal position (1/4,1/4,1/8). In particular, the coordinate *x* of Se change from 0.264 at ambient pressure to 0.257 at 13.3 GPa (the highest pressure were we observed a *pure* defect-stannite structure), while the *z* coordinate of Se changes from 0.117 to 0.121. As a consequence of all these changes on the anion coordinates, the Ga-Se bond corresponding to the Ga atom located at the position 2a is reduced a 20% more than the other cation-Se bonds. Additionally, all the cation-Se bonds increase approximately a 3%; and the Se-Se distances decrease a 3% at the phase transition. The second change is caused by the volume collapse of the structure of $ZnGa_2Se_4$, while the first one is caused by the reordering of the cation positions. We also found that upon compression $CdGa_2S_4$ becomes more symmetric and that the position of the S atoms gradually approach the ideal position (1/4,1/4,1/8). In particular the *x* coordinate of S changes from 0.270 at ambient pressure to 0.265 at 17 GPa and the *y* coordinate of S changes from 0.260 to 0.250. On the other hand, the *z* coordinate of S changes from 0.140 to 0.130. As a consequence of all these changes on the anion coordinates, the Cd-S bond is reduced around 50% more than the other cation-Se bonds. A similar preferred compressibility of the Cd-Se bonds has been reported in $CdGa_2Se_4$ [7]. Additionally, all the cation-S bonds increase approximately 8% and the S-S distances decrease 8% at the transition. As in $ZnGa_2Se_4$, the second change in $CdGa_2S_4$ is caused by a volume collapse and the first one by reordering of the cation positions.

It is interesting to see that in $ZnGa_2Se_4$ and $CdGa_2Se_4$ the Ga-Se bonds have a similar average compressibility. The same behavior can be deduced for $MnGa_2Se_4$ from the data reported in Ref. [8]. Since in ternary compounds the bulk compressibility is related to the polyhedral compressibility [29, 30] it is not strange that the three studied digallium tetraselenides have a similar compressibility. Therefore a bulk modulus close to 45 GPa should be expected also for $HgGa_2Se_4$ [31].

We mentioned above that an increase of the cationic disorder apparently takes place in $ZnGa_2Se_4$. This disorder may be the origin of the precursor effects of the transition observed in x-ray experiments in $CdGa_2Se_4$ [7] and in Raman experiments in four different digallium tetraselenides and tetrasulphides [3]. Optical-absorption measurements also detect the precursor effects on $ZnGa_2Se_4$ and $CdGa_2Se_4$ around 13 GPa, which are responsible of non-reversible changes on the optical-absorption edge [27]. Earlier it was shown that order-disorder phase transition takes place in defect-chalcopyrite tetraselenides at high temperatures and ambient pressure only when the tetragonal distortion parameter $\delta = 2 - c/a$, is smaller than 0.05 [32]. A considerable reduction of this parameter was observed both in $MnGa_2Se_4$ and $CdGa_2Se_4$ before the transition to the rock-salt structure [7, 8]. In our case, we observed that this parameter decreases from 0.165 at ambient pressure to 0.087 at 17 GPa for $CdGa_2Se_4$. In $ZnGa_2Se_4$ $\delta$ is equal to 0.022 at ambient pressure and approaches zero before the phase transition. On top of that, we also found in both compounds a decrease of the anion distortion parameter $\sigma_0 = \sqrt{(x-0.25)^2 + (y-0.25)^2 + (z-0.125)^2}$ upon compression.

This parameter decreases for $ZnGa_2Se_4$ from 0.0214 at ambient pressure to 0.0107 at 13.3 GPa and from 0.117 at ambient pressure to 0.052 at 17 GPa for $CdGa_2S_4$. A similar decrease was also found upon compression for $MnGa_2Se_4$ [8]. According with Garbato *et al.* [19], both the reduction of $\delta$ and $\sigma_0$ causes an increase of the cation disorder. Thus, the relation suggested above is fully consistent with the Raman and diffraction peak broadening and the increase of the crystalline disorder observed during compression.

**III.D. Size criterion**

According to an empirical rule (size criterion) proposed by Jayaraman *et al.* [33] the transition pressure from tetrahedral to octahedral coordination in $ABX_2$ compounds (e.g. chalcopyrite $CuGaSe_2$) increases with decreasing the ratio between the averaged cation radius, $(r_A + r_B)/2$, and the anion radius, $r_X$. Other size criteria, similar to Jayaraman's rule, have been proven to work satisfactory to predict the transition pressure in ternary compounds [34, 35]. However, Beister *et al.* [36] have challenged Jayaraman´s rule based upon data on $LiInSe_2$, $CuInSe_2$ and $AgInSe_2$. These authors proposed the transition pressure should increase with the decreasing cation radius difference $|r_A - r_B|$. Other authors tried to rationalize the transition pressure from a fourfold-coordinated structure to a sixfold-coordinated structure using the crystal ionicity [3], but this approach did not give a clear systematic for the transition pressures of $ABX_2$ and $AB_2X_4$ compounds. By comparing all the data available on the literature on twenty-three different tetragonal-coordinated $ABX_2$ and $AB_2X_4$ compounds, we will show that the transition pressure on these compounds can be rationalized using the ionic radius on cations and anions in a similar way than proposed by Jayaraman. Table III displays the transition pressures for twenty-three different $ABX_2$ and $AB_2X_4$ compounds with structures related to sphalerite and the ionic radii of the elements A, B, and X [37]. Fig. 6 shows the transition pressure as a function of $\xi = (r_A + r_B)/2r_X$. In the figure, it can be seen that most of the compounds of interest follows two clear systematics. One for those compounds with large cations ($1.35 > r_A + r_B > 1.15$) and another for the compounds with small cations ($r_A + r_B < 1.15$). It is interesting to see that also spinel-structured $AB_2X_4$ compounds seem to follow the same systematic. In particular, compounds like $ZnAl_2S_4$, $CuIr_2S_4$, $MgAl_2O_4$, $CuCrZrS_4$, and $Zn_2TiO_4$ match very well with the systematic reported in Fig. 6 (see Table III). As shown in Fig. 6, apparently also the compounds with $1.45 > r_A + r_B > 1.35$ and $1.45 > r_A + r_B$ have a similar tendency to the increase of the transition pressure with the increase of ξ. It is important to note here that the fitting lines shown in Fig. 6 for each cation-size regime are dependent on the high-pressure points. Unfortunately, in the literature there are few data available for ξ values from 0.35 to 0.4, which make the systematic here proposed only valid to make back-of-the-envelope estimations for transition pressures in compounds not studied upon compression yet. Regarding the criterion proposed by Beister *et al.*, it is clear from Table III that no possible correlation could be established between the transition pressures and cation radius difference $|r_A - r_B|$. We strongly believe that the comparison made by Beister *et al.* between $LiInSe_2$, $CuInSe_2$ and $AgInSe_2$ was inadequate and it also mislead these authors to challenge Jayaraman's rule. $LiInSe_2$ has an orthorhombic structure (*Pna2₁*) which is not related to sphalerite and furthermore its cations do not have a tetrahedral coordination. Therefore, phenomenological comparisons are not possible between $LiInSe_2$ and the other two compounds. Further, the transition pressure for $AgInSe_2$ (2.5 GPa) was taken by Beister *et al.* from resistivity and x-ray diffraction measurements performed under highly non-hydrostatic conditions [38]. Uniaxial stresses are known to strongly affect pressure-induced phase transitions reducing the transition pressure by 10 GPa [11]. On the other hand, selenides usually have higher transition pressures than isomorphous tellurides, and $AgInTe_2$ is known to remain stable at least up to 2.7 GPa according to x-ray diffraction experiments [39]. So the early studies on $AgInSe_2$ are not good candidates to establish a systematic for $ABX_2$ compounds. From the systematic behavior of Fig. 6, a transition pressure of 8(2) GPa is predicted for $AgInSe_2$ ($r_A + r_B = 1.60$ and $\xi = 0.40$). This suggests that studies taking advantage of the state-of-the-art synchrotron facilities are needed to clarify the high-pressure structural behavior of $AgInSe_2$. It would be also interesting to perform such studies on $AgInS_2$ and $AgInTe_2$. For these compounds we predict transition pressures of 12(2) GPa and 4(2) GPa, respectively. In $AgInS_2$ and $AgInTe_2$ x-ray diffraction experiments have been performed only up to 5 GPa [40] and 2.7 GPa [39], respectively, and the chalcopyrite structure found to be stable up to these pressures. An extension of these studies is required to test our predictions. The systematic proposed in this work could be also applied to other sphalerite-related compounds like defect-chalcopyrite $HgGa_2Se_4$ [31] and double-defective chalcopyrite $InPS_4$ [41], for which transition pressures of 9(2) and 7(2) GPa are predicted. It is important to note here, that the systematic established in this work cannot be applied to $ABX_2$ and $AB_2X_4$ compounds with structures not related to sphalerite similar to $LiInSe_2$, and like most of the alkaline-earth digallium tetraselenides, which usually crystallize in an orthorhombic structure [42]. However, the same systematic apparently works well in spinel-structured $AB_2X_4$ compounds. At least in those where the structural stability of the compound is not affected by Jahn-Teller effects caused by the presence of magnetic ions. One example of these compounds is $MgAl_2O_4$ (see Table III and Fig. 6). Based upon these facts, predictions can be made for the transition pressures of $MgGa_2O_4$ and $MgIn_2O_4$, for which transition pressures of 28(3) GPa and 26(3) GPa are predicted. For the cases of $ZnAl_2O_4$ and $CdAl_2O_4$ we have predicted transition pressures as 37(4) GPa and 33(4) GPa respectively.

To conclude the discussion, we would like to comment that in $CuGaTe_2$ and $CuInTe_2$ the following pressure-induced structural sequence is observed: chalcopyrite → rock-salt → *Cmcm* [43]. The same structural sequence has been observed in sphalerite-structured binary semiconductors like CdTe [25]. Therefore, upon cystallochemical arguments [26] it is quite reasonable to speculate the sphalerite-related to NaCl-type transition for all the compounds reviewed here. Our results further suggest that at higher pressures a second transition to a *Cmcm* structure could take place in most of them.

**IV. Conclusions**

High-pressure ADXRD experiments on two sphalerite-related defective semiconductors have been reported. The results obtained show that $ZnGa_2Se_4$ has a tetragonal defect-stannite structure from atmospheric pressure to 15.2 GPa. Furthermore, the tetragonal structure co-exists with a higher-symmetry cubic structure from 15.2 to 18.5 GPa, and beyond this pressure, only the cubic structure is stable up to 23 GPa. In $CdGa_2S_4$ it has been found that the structure is a tetragonal defect chalcopyrite up to 17 GPa and beyond a pressure-induced phase transition takes place to a cubic structure similar to $ZnGa_2Se_4$. The results obtained were compared with those previously reported in isostructural compounds. The role played by cation disorder in the observed transition is discussed. Finally, a room temperature equation EOS for the title compounds is reported and a systematic for pressure-driven phase transitions in sphalerite-related compounds is discussed.

**Acknowledgments**

This study was supported by the Spanish government MEC under Grants No: MAT2007-65990-C03-01, MAT2006-02279, and CSD-2007-00045 and the Generalitat Valenciana (Project GV06/151). The U.S. Department of Energy, Office of Science, and Office of Basic Energy Sciences supported the use of the APS under Contract No. W-31-109-Eng-38. DOEBES, DOE-NNSA, NSF, DOD-TACOM, and the Keck Foundation supported the use of the HPCAT. D. Errandonea is indebted to the Fundación de las Artes y las Ciencias de Valencia for granting him the IDEA prize. F.J. Manjón acknowledges also financial support from "Vicerrectorado de Innovación y Desarrollo de la UPV" through project UPV2008-0020. Work at UNLV is supported by DOE award No. DEFG36-05GO08502. The UNLV High Pressure Science and Engineering Center is supported by the DOENNSA under cooperative agreement No. DE-FC52-06NA26274.

**Table I:** Atomic positions and site occupation fraction (SOF) refined for $ZnGa_2Se_4$ at ambient pressure. Defect-stannite structure, space group: $I\bar{4}2m$, Z =2.

| Atom | Site | *x* | *y* | *z* | SOF |
|---|---|---|---|---|---|
| Ga | 2a | 0 | 0 | 0 | 1 |
| Ga | 4d | 0 | 1/2 | 1/4 | 0.5 |
| Zn | 4d | 0 | 1/2 | 1/4 | 0.5 |
| Vacancy | 2b | 0 | 0 | 1/2 | 0 |
| Se | 8i | 0.264(8) | 0.264(8) | 0.117(4) | 1 |

**Table II:** Atomic positions and site occupation fraction (SOF) refined for $CdGa_2S_4$ at ambient pressure. Defect-chalcopyrite structure, space group: $I\bar{4}$, Z =2.

| Atom | Site | *x* | *y* | *z* | SOF |
|---|---|---|---|---|---|
| Cd | 2a | 0 | 0 | 0 | 1 |
| Ga | 2b | 0 | 0 | 1/2 | 1 |
| Ga | 2c | 0 | 1/2 | 1/4 | 1 |
| Vacancy | 2d | 0 | 1/2 | 3/4 | 0 |
| S | 8i | 0.271(8) | 0.261(8) | 0.140(4) | 1 |

**Table III:** Transition pressures and ionic radii of different sphalerite-related semiconductors. The factor $\xi = (r_A + r_B)/2\ r_X$ was calculated for each compound using the Shannon radii [37].

| Compound | Structure | $r_A$ | $r_B$ | $r_X$ | $\xi$ | $\lvert r_A - r_B \rvert$ | $P_T$(GPa) | Reference |
|---|---|---|---|---|---|---|---|---|
| $CuAlTe_2$ | Chalcopyirite | 0.60 | 0.39 | 2.21 | 0.2239 | 0.21 | 8.3 | 44 |
| $CuGaTe_2$ | Chalcopyirite | 0.60 | 0.47 | 2.21 | 0.2421 | 0.13 | 9.4 | 45 |
| $CuAlSe_2$ | Chalcopyirite | 0.60 | 0.39 | 1.98 | 0.2500 | 0.21 | 13.2(1.2) | 44, 46 |
| $ZnAl_2Se_4$ | Defect Chalcopyirite | 0.60 | 0.39 | 1.98 | 0.2500 | 0.21 | 14.4 | 47 |
| $CuAlS_2$ | Chalcopyirite | 0.60 | 0.39 | 1.84 | 0.2690 | 0.21 | 16.5(1.5) | 44, 46 |
| $CuGaSe_2$ | Chalcopyirite | 0.60 | 0.47 | 1.98 | 0.2702 | 0.13 | 13.6(0.5) | 3 |
| $ZnGa_2Se_4$ | Defect Stannite | 0.60 | 0.47 | 1.98 | 0.2702 | 0.13 | 16.5(2) | 3, 5 |
| $ZnAl_{1.2}Ga_{0.8}S_4$ | Chalcopyrite | 0.60 | 0.42 | 1.84 | 0.2771 | 0.18 | 17 | 48 |
| CuGaSSe | Chalcopyrite | 0.60 | 0.47 | 1.91 | 0.2801 | 0.13 | 17(1) | 49 |
| $MnGa_2Se_4$ | Defect Chalcopyirite | 0.66 | 0.47 | 1.98 | 0.2854 | 0.19 | 13(1) | 8 |
| $CuGaS_2$ | Chalcopyirite | 0.6 | 0.47 | 1.84 | 0.2907 | 0.13 | 16.5(0.5) | 3 |
| $ZnGa_2S_4$ | Defect Stannite | 0.60 | 0.47 | 1.84 | 0.2907 | 0.13 | 21(1) | 3 |
| $ZnAl_2S_4$ | Spinel | 0.60 | 0.53 | 1.84 | 0.3070 | 0.07 | 20(2) | 50 |
| $MgAl_2O_4$ | Spinel | 0.57 | 0.53 | 1.38 | 0.3985 | 0.04 | 35(5) | 51, 52 |
| $CuInTe_2$ | Chalcopyirite | 0.60 | 0.62 | 2.21 | 0.2692 | 0.02 | 4.2(1.4) | 45 |
| $CdAl_2Se_4$ | Defect Chalcopyirite | 0.78 | 0.39 | 1.98 | 0.2955 | 0.39 | 9.1(1) | 47 |
| $CuInSe_2$ | Chalcopyirite | 0.60 | 0.62 | 1.98 | 0.3081 | 0.02 | 7.1 | 53 |
| $CdGa_2Se_4$ | Defect Chalcopyirite | 0.78 | 0.47 | 1.98 | 0.3156 | 0.31 | 16(1.5) | 3, 6, 7 |
| $CdAl_2S_4$ | Defect Chalcopyirite | 0.78 | 0.39 | 1.84 | 0.3179 | 0.39 | 14(0.5) | 47, 54 |
| $CuInS_2$ | Chalcopyirite | 0.60 | 0.62 | 1.84 | 0.3315 | 0.02 | 9.6 | 3 |
| $CuIr_2S_4$ | Spinel | 0.57 | 0.68 | 1.84 | 0.3396 | 0.11 | 13(1) | 55 |
| $CdGa_2S_4$ | Defect Chalcopyirite | 0.78 | 0.47 | 1.84 | 0.3396 | 0.31 | 17(2) | 3, 4 |
| $HgAl_2Se_4$ | Defect Chalcopyirite | 0.96 | 0.39 | 1.98 | 0.3409 | 0.57 | 12.7 | 47 |
| $AgGaS_2$ | Chalcopyirite | 0.79 | 0.47 | 1.84 | 0.3424 | 0.32 | 15 | 3 |
| $CuCrZrS_4$ | Spinel | 0.57 | 0.72 | 1.84 | 0.3505 | 0.15 | 15 | 56 |
| $Zn_2TiO_4$ | Inverse Spinel | 0.60 | 0.60 | 1.38 | 0.4347 | 0 | 28(3) | 57 |
| $AgGaTe_2$ | Chalcopyirite | 0.98 | 0.47 | 2.21 | 0.3320 | 0.51 | 5.4 | 58 |
| $CdCr_2Se_4$ | Spinel | 0.78 | 0.62 | 1.98 | 0.3535 | 0.16 | 9(1) | 59 |
| $AgGaSe_2$ | Chalcopyirite | 0.98 | 0.47 | 1.98 | 0.3661 | 0.51 | 8.3 | 3 |
| $MgIn_2S_4$ | Spinel | 0.57 | 0.80 | 1.84 | 0.3723 | 0.23 | 10.5(1.5) | 60 |
| $MnIn_2S_4$ | Spinel | 0.66 | 0.80 | 1.84 | 0.3967 | 0.14 | 8(1) | 60 |
| $CdIn_2S_4$ | Spinel | 0.78 | 0.80 | 1.84 | 0.4293 | 0. 02 | 11(1) | 60 |

**Figure captions**

**Figure 1: (color online)** (a) Defect-chalcopyrite structure of $CdGa_2S_4$. Blue circles: S, red circles: Cd, and magenta circles: Ga. (b) Defect-stanniite structure of $ZnGa_2Se_4$. Blue circles: Se, red circles: Ga, and magenta circles: Ga and Zn atoms (SOF = 0.5 for each one).

**Figure 2:** X-ray diffraction patterns of $ZnGa_2Se_4$ at selected pressures. The background was subtracted. At 0.87 and 23 GPa the collected pattern (dots) is shown together with the refined patters (solid line) and the residuals of the refinement. These two patterns illustrate the quality of the refinements obtained for the low- and high-pressure phases. The asterisks shows the appearance of the peaks of the high-pressure phase.

**Figure 3:** X-ray diffraction patterns of $CdGa_2S_4$ at selected pressures. The background was subtracted. At 0.6 and 22 GPa the collected pattern (dots) is shown together with the refined patters (solid line) and the residuals of the refinement. These two patterns illustrate the quality of the refinements obtained for the low- and high-pressure phases.

**Figure 4:** Pressure evolution of the volume and the lattice parameters of $ZnGa_2Se_4$. Solid squares: low-pressure phase. Solid circles: high-pressure phase. Empty circles: high-pressure phase on pressure release. The solid lines represents the reported EOS and for the lattice parameters are a just guide to the eye.

**Figure 5:** Pressure evolution of the volume and the lattice parameters of the low-pressure phase of $CdGa_2S_4$. Solid squares: low-pressure phase. Solid circles: high-pressure phase. The solid lines represents the reported EOS and for the lattice parameters are a just guide to the eye.

**Figure 6:** Transition pressures of different $ABX_2$ and $AB_2X_4$ compounds as a function of $\xi = (r_A+r_B)/2\ r_X$. The solid lines estimates the transition pressure of different

compounds the dotted lines gives the deviation from the estimated value. The data plotted corresponds to those shown in Table III.

**Figure 1**

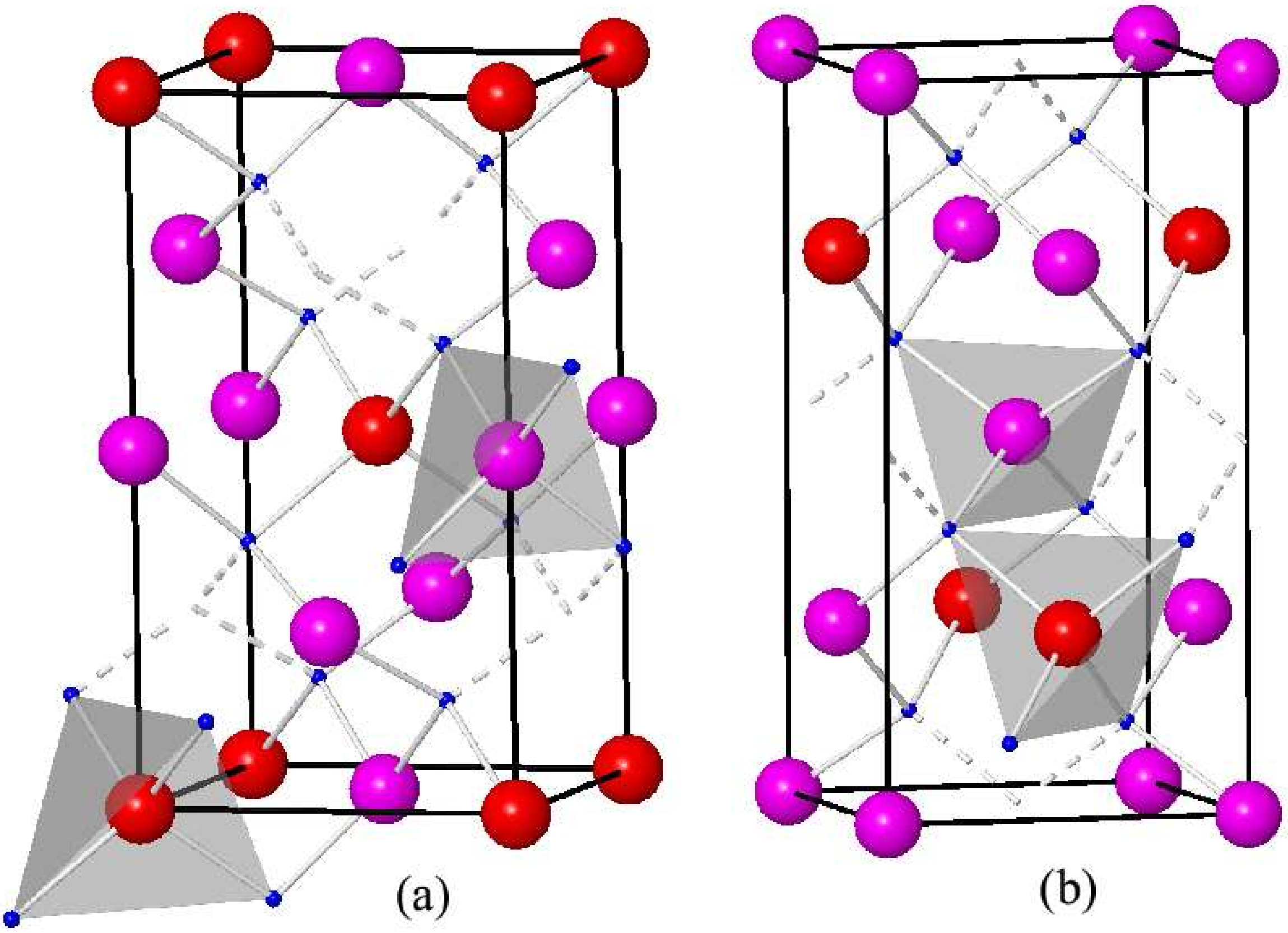

**Figure 2**

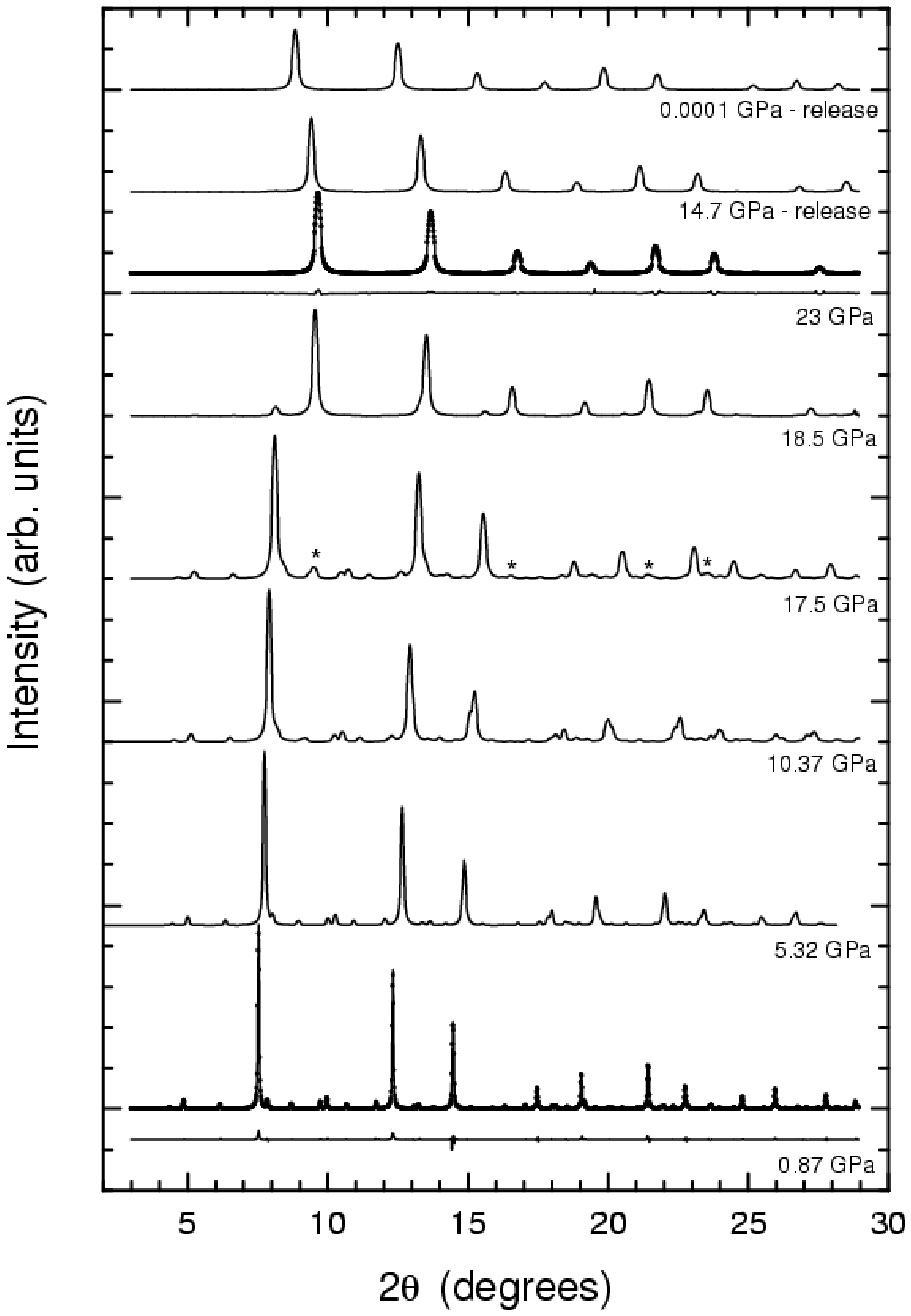

**Figure 3**

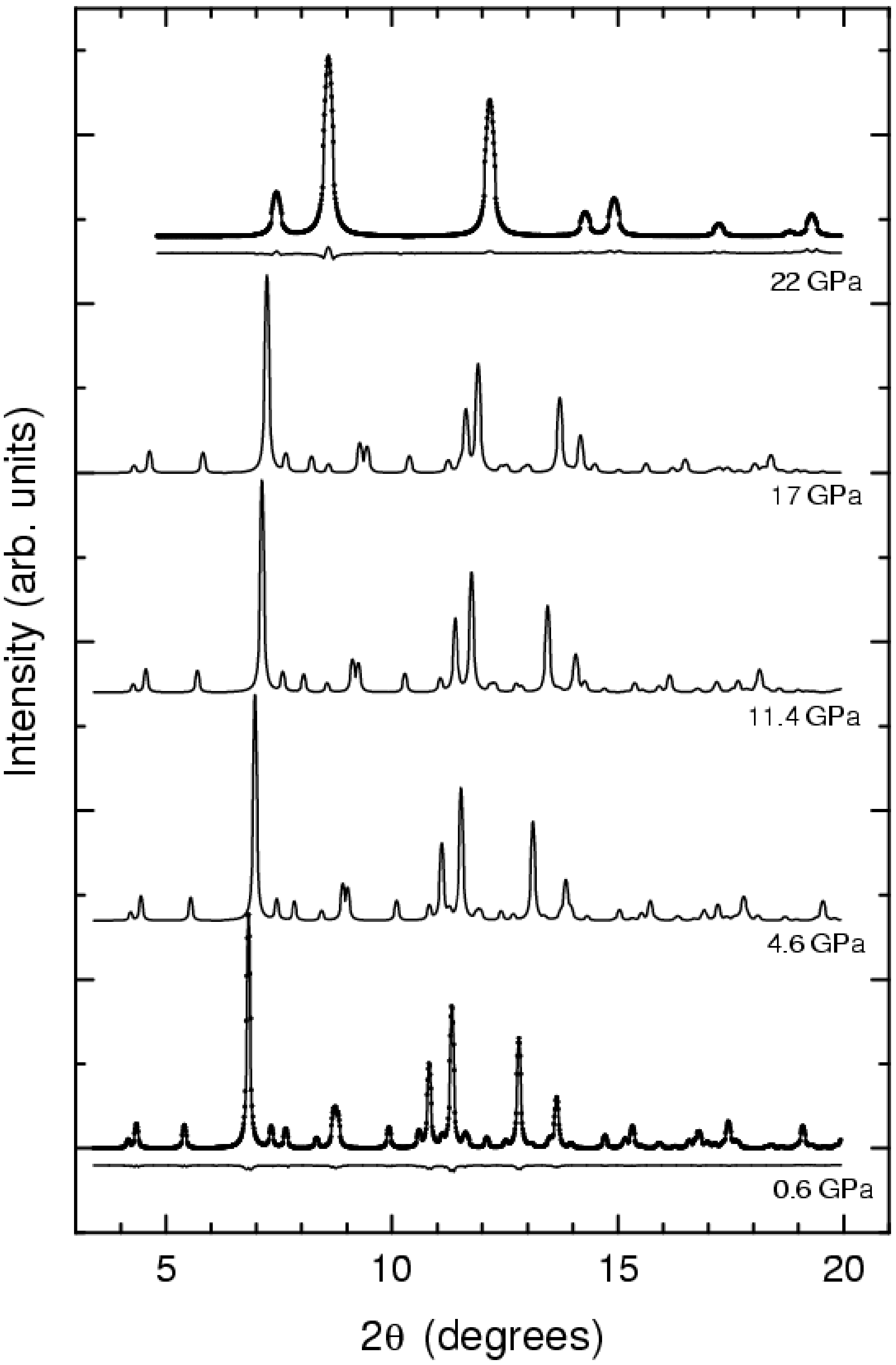

Figure 4

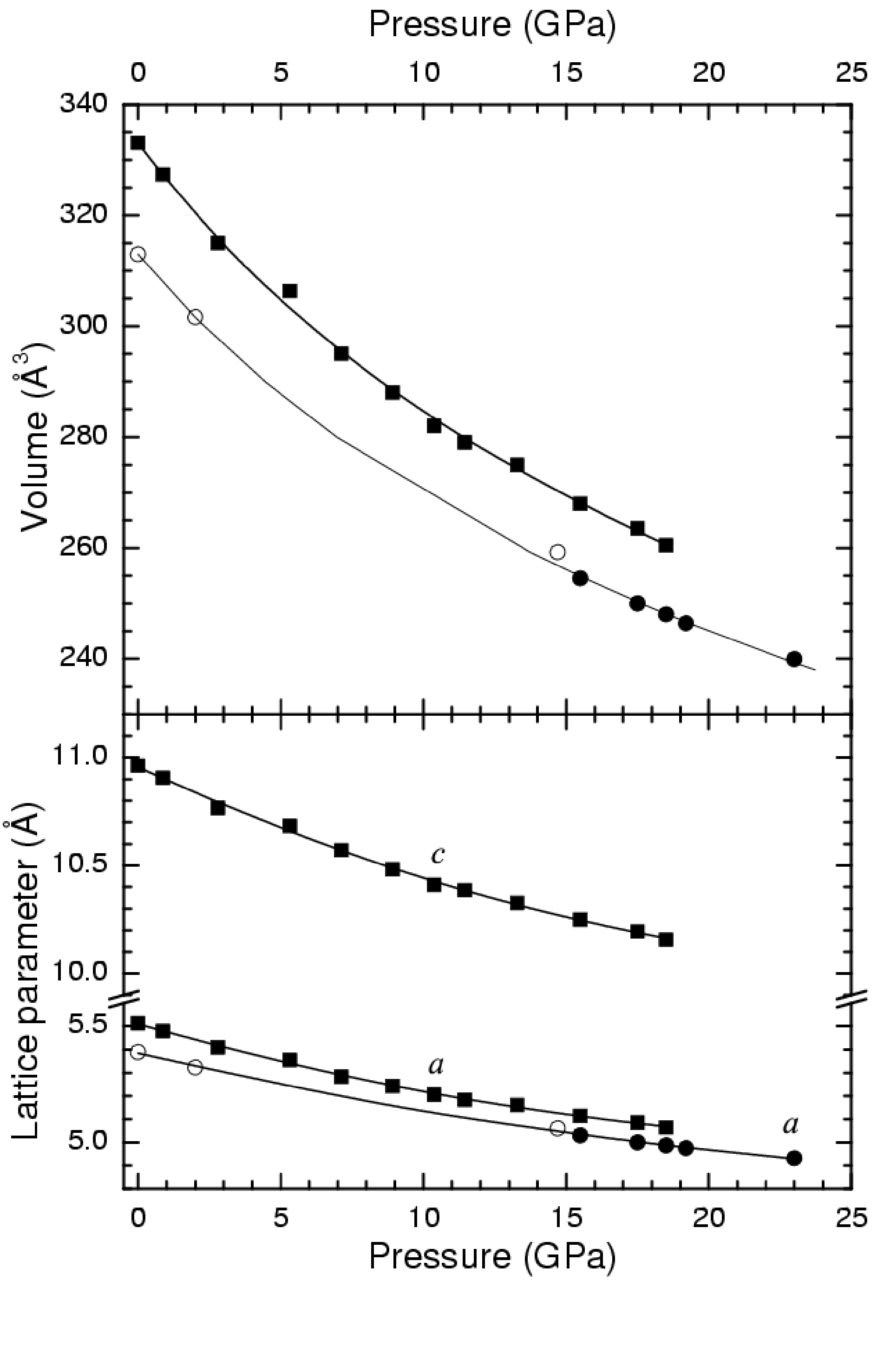

**Figure 5**

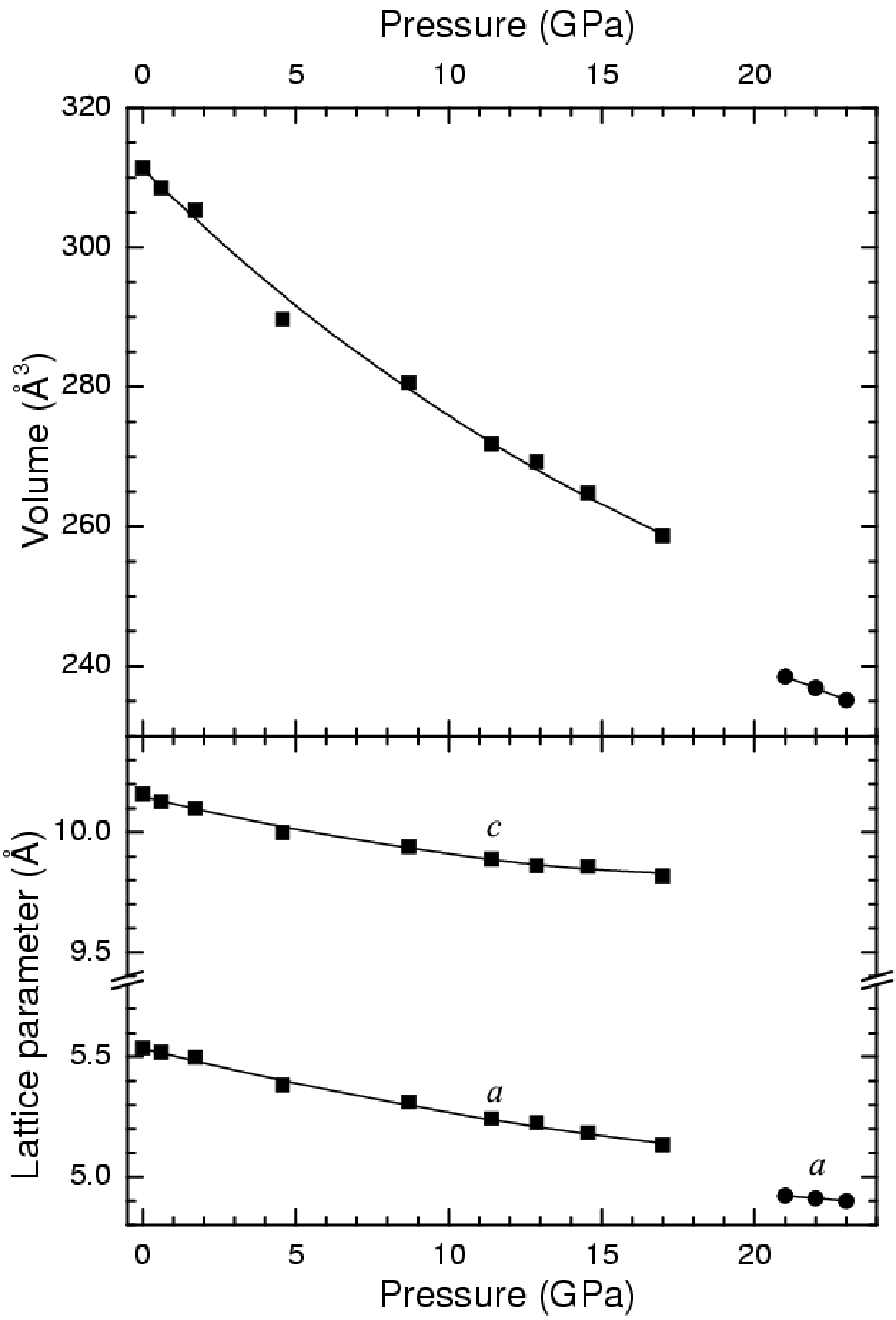

**Figure 6**

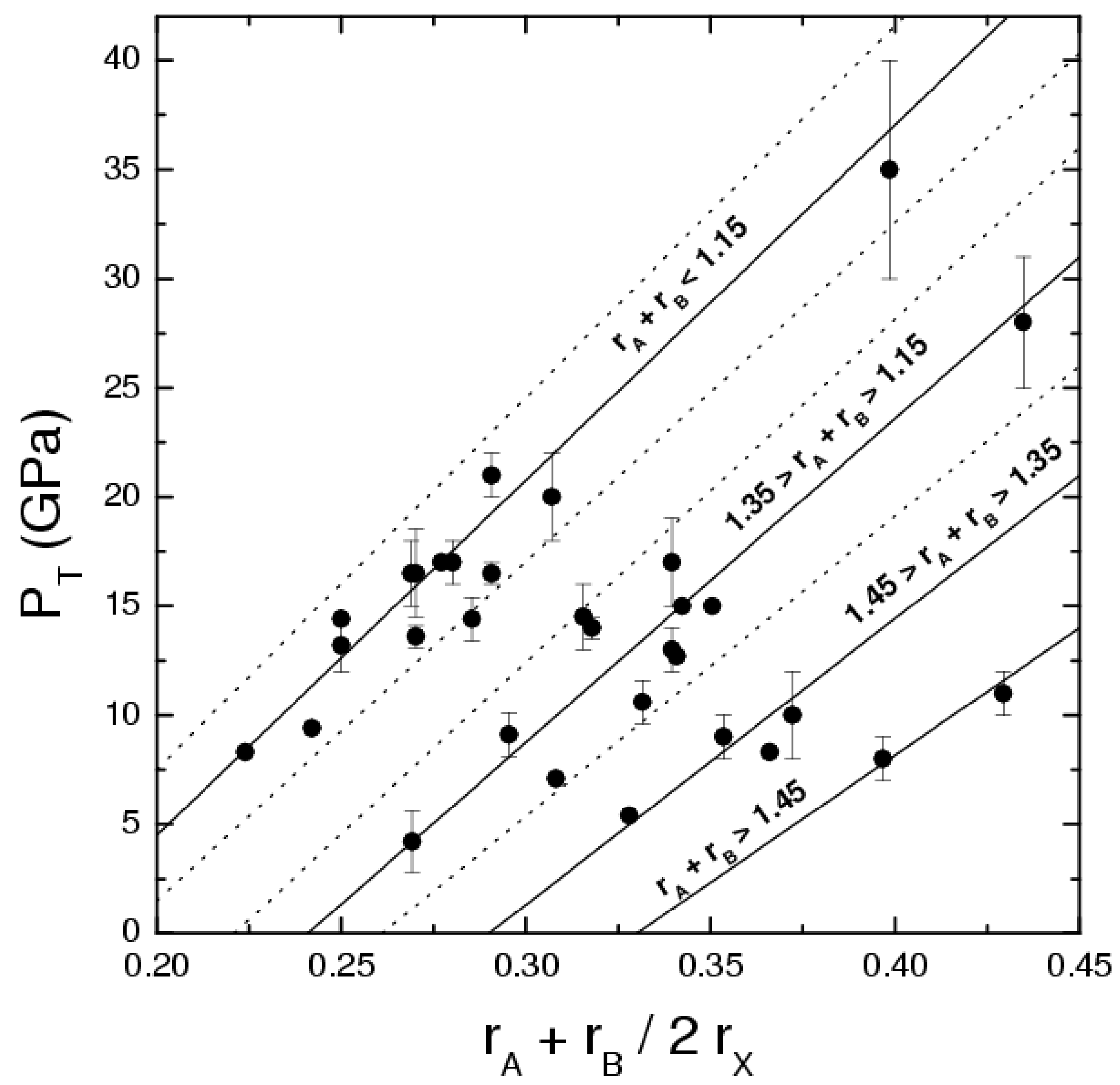

$P_T$ (GPa)
$r_A + r_B / 2\ r_X$
$r_A + r_B < 1.15$
$1.35 > r_A + r_B > 1.15$
$1.45 > r_A + r_B > 1.35$
$r_A + r_B > 1.45$
0
5
10
15
20
25
30
35
40
0.20
0.25
0.30
0.35
0.40
0.45